\documentclass{desyproc}
\newcommand*{\ee}{e^+e^-}

\begin{document}
%------------------------------------
\title{\begin{center}Non-linear quantum dynamics
in strong and short electromagnetic fields\end{center}}

%for single authors the superscripts are optional
\author{{\slshape
  Alexander~I.~Titov$^{1}$,   Burkhard~K\"ampfer$^{2,3}$, A Hosaka$^4$ and H. Takabe$^2$ }\\[1ex]
 $^1$Bogoliubov Laboratory of Theoretical Physics, JINR, Dubna 141980, Russia\\
 $^2$Helmholtz-Zentrum  Dresden-Rossendorf, 01314 Dresden, Germany\\
 $^3$Institut f\"ur Theoretische Physik, TU~Dresden, 01062 Dresden, Germany\\
 $^4$RCNP, 10-1 Mihogaoka Ibaraki, 567-0047 Osaka, Japan}

% please enter the contribution ID for the DOI
\contribID{10.3204/DESY-PROC-2016-04}

% TO THE CONFERENCE EDITORS:
% please update the following information
% before sending the template to the authors
\confID{999}  % if the conference is on Indico uncomment this line
\desyproc{DESY-PROC-2016-04}
\acronym{VIP2010} % if you want the Acronym in the page footer uncomment this line
\doi  % if there is an online version we will register DOIs

\maketitle

\begin{abstract}
 In our contribution we give a brief overview
 of two widely discussed quantum processes:
 electron-positron pairs production off
 a probe photon propagating through
 a polarized short-pulsed electromagnetic (e.m.) (e.g.\ laser)
 wave field or generalized Breit-Wheeler process
 and a single a photon emission off an electron interacting
 with the laser pules, so-called non-linear Compton scattering.
 We show that at small and moderate laser field intensities
 the shape and duration of the pulse are very important
 for the probability of considered processes. However,
 at high intensities the multi-photon interactions
 of the fermions with laser field
 are decisive and completely determined all aspects of subthreshold
 $\ee$ pairs and photon production.
 \end{abstract}

\section{Introduction}

 The rapidly progressing laser technology \cite{Tajima}
 offers unprecedented opportunities for investigations
 of quantum systems with intense laser beams~\cite{Piazza}.
 A laser intensity $I_L$ of $\sim 2\times 10^{22}$  W/cm${}^2$ has been already
 achieved~\cite{I-22}. Intensities of the order of
 $I_L \sim 10^{23}...10^{25}$ W/cm$^2$ are envisaged in near future, e.g.\
 at the CLF~\cite{CLF}, ELI~\cite{ELI}, HiPER~\cite{hiper}.
 Further facilities are in planning on construction stage, e.g.
 PEARL laser facility~\cite{sarov} at Sarov/Nizhny Novgorod, Russia.
 The high intensities are provided in short
 pulses on a femtosecond pulse duration
 level~\cite{Piazza,ShortPulse,ShortPulse_2},
 with only a few oscillations of the electromagnetic (e.m.) field
 or even sub-cycle pulses.
 (The tight connection of high intensity and short pulse duration
 is further emphasized in \cite{Mackenroth-2011}. The attosecond
 regime will become accessible at shorter wavelengths~\cite{atto,I-222}).

 Quantum processes occurring in the interactions  of charge fermions
 in very (infinitely) long e.m. pulse  were investigated in detail
 in the pioneering works of
 Reiss~\cite{Reiss} as well as Narozhny, Nikishov
 and Ritus~\cite{NikishovRitus,NR-64,Ritus-79}.
 We call the such approaches as an infinite pulse approximation (IPA)
 since it refers to a stationary scattering process.
 Many simple and clear expressions for the production probabilities
 and cross sections have been obtain within IPA. It was shown
 that the charged fermion (electron, for instance)
 can interact with $n\ge1$ photon simultaneously ($n$ is an integer number),

 However, recently it has become clear
 that for the photon production
 off an electron interacting with short laser pulse (Compton scattering)
 and for $\ee$ pair production off a probe photon interacting with
 short e.m. pulses (Breit-Wheeler process)
 the finite pulse shape and the pulse duration become important
 (see, for example~\cite{TitovPEPAN} and reference their in).
 That means the treatment of the intense and short laser field as an infinitely
 long wave train is no longer adequate. The theory must
 operate with essentially finite pulse. We call
 such approaches as a finite pulse approximation (FPA).

 In this contribution we consider some particularities
 of generalized Breit-Wheeler and Compton processes in a
 short and strong laser pulses. For this purpose
 we use the widely employed
 the four electromagnetic (e.m.) potential
 for a circularly polarized laser field in
 the axial gauge $A^\mu=(0,\,\mathbf{A}(\phi))$ with
\begin{eqnarray}
 \mathbf{A}(\phi)=f(\phi) \left( \mathbf{a}_1\cos(\phi+\tilde\phi)+ \mathbf
 {a}_2\sin(\phi+\tilde\phi)\right)~, \label{III1}
 \end{eqnarray}
 where $\phi=k\cdot x$ is invariant phase with four-wave vector
 $k=(\omega, \mathbf{k})$, obeying the null field property $k^2=k\cdot
 k=0$ (a dot between four-vectors indicates the Lorentz scalar
 product) implying $\omega = \vert\mathbf{k}\vert$,
 $ \mathbf{a}_{(1,2)} \equiv \mathbf{a}_{(x,y)}$;
 $|\mathbf{a}_x|^2=|\mathbf{a}_y|^2 = a^2$, $\mathbf{a}_x \mathbf{a}_y=0$;
 transversality means $\mathbf{k} \mathbf{a}_{x,y}=0$ in the present gauge.
 The envelope function $f(\phi)$ with
 $\lim\limits_{\phi\to\pm\infty}f(\phi)=0$ accounts for the
 finite pulse length.
 We are going to analyze dependence of observables on
 the shape of $f(\phi)$ in Eq.~(\ref{III1}) for two types of envelopes:
 the one-parameter hyperbolic secant~(hs) shape and the two-parameter
 symmetrized Fermi~(sF) shape widely used for parametrization of the
 nuclear density~\cite{Luk}:
 $f_{\rm hs}(\phi)=(\cosh\phi/{\Delta})^{-1}$ and
 $f_{\rm sF}(\phi)=(\cosh{\Delta}/{b} +1)
 (\cosh{\Delta}/{b} +\cosh{{\phi}/{b}})^{-1}$.
 The parameter $\Delta$ characterizes the
 pulse duration $2\Delta$ with $\Delta=\pi N$,
 where $N$ has a meaning
 of a "number of oscillations" \ in the pulse.
 The parameter $b$ in the sF shape describes the ramping time
 in the neighborhood of $\phi \sim \Delta$. Small values of ratio $b/\Delta$
 cause a flat-top shaping. At $b/\Delta\to0$, the sF shape
 becomes a rectangular pulse.
 In the following, we choose the ratio
 $b/\Delta$ as the second independent
 parameter for the sF envelope function.
 %%%%%%%%%%%%%%%%%%%%%%%%%%%%%%%%%%%%%%%%%%%%%%%%%%%%%%%%%%%%%%%%%%%%%%
 These two shapes cover
 a variety of relevant envelopes discussed in literature
 (for details see~\cite{TitovPRA}).
 %%%%%%%%%%%%%%%%%%%%%%%%%%%%%%%%%%%%%%%%%%%%%%%%%%%%%%%%%%%%%%%%%%%%%%%%%%%%%%%
  The carrier envelope phase $\tilde\phi$
  is particularly important for the short the pulse duration with $N\leq1$.
  Therefore we start our presentation with case of
  $\tilde\phi=0$ and discuss impact of finite carrier phase at the end.
  Finally we note that, the interaction of the background field is
  determined by dimensionless reduced e.m. intensity $\xi^2=\sqrt{-A^2}/M_e^2$,
  where $M_e$ is the electron mass (we use natural units with
 $c=\hbar=1$, $e^2/4\pi = \alpha \approx 1/137.036$). (for more detail see~\cite{TitovPEPAN}).
%%%%%%%%%%%%%%%%%%%%%%%%%%%%%%%%%%%%%%%%%%%%%%%%%%%%%%%%%%%%%%%%%%%%%%%%%%%%%%%%

 Some important difference between IPA and FPA is
 that in the first case the variable
 $n = 1, 2,\, \cdots$ is integer, it refers to
 the contribution of the individual harmonics.
 The value $n\omega$ is related to the energy
 of the background field involved into considered
 quantum process. Obviously, this value is a multiple of $\omega$.
 In FPA, the basic subprocess operate with $l$
 {\it background photons},
 where $l$ is a continuous variable.
 The quantity $l\omega$ can be considered as the energy
 partition of the laser beam involved into considered process,
 and it is not a multiple $\omega$.
 Mindful of this fact, without loss of generality, we
 denote the processes with $l>1$ as a generalized multi-photon
 processes, remembering that $l$ is a continuous quantity.
%%%%%%%%%%%%%%%%%%%%%%%%%%%%%%%%%%%%%%%%%%%%%%%%%%%%%%%%%%%%%%%%%%\

 This lecture is based on the review paper~\cite{TitovPEPAN}
 and is organized as follows.
 Sect.~2 is devoted to the non-linear Breit-Wheeler
 process.
 In Sect.~3 we discuss several aspects of non-linear Compton scattering
 for short and sub-cycle pulses.
 Our conclusions are presented in Sect.~5.

%%%%%%%%%%%%%%%%%%%%%%%%%%%%%
\section{The ${\mathbf{\ee}}$ pair production in a finite pulse}
%%%%%%%%%%%%%%%%%%%%%%%%%%%%%

 We consider $\ee$ pair production in the interaction of a probe photon with
 a circularly polarized e.m.field~(\ref{III1}
 within the Furry picture, which diagrammatically is
 represented by a one-vertex graph, describing the decay
 of the probe photon with the four-momentum $k'$ into a laser dressed
 $\ee$ pair. The presence of the background e.m. field
 is included in the Volkov solution of the outgoing
 $e^+$ and $e^-$. (In the weak-field approximation this graph
 turns into the known two two-vertex graphs for the perturbative
 Breit-Wheeler process).
 Contrary to the IPA, utilization of~(\ref{III1})
 the Volkov solutions in FPA  assume
 all fermion momenta and masses
 take their vacuum values  $p$ and $m$, respectively,
 whereas the corresponding wave functions are modified
 in accordance with the Volkov solution~\cite{Volkov,LL4}
 (with more complicated compare to IPA, phase factor).
The finite (in space-time) e.m. potential (\ref{III1}) for FPA
requires the use of Fourier integrals
for invariant amplitudes, instead of Fourier
series which are employed in IPA.
The partial harmonics
become thus continuously in FPA.
The $S$ matrix element is expressed generically as
\begin{eqnarray}
S_{fi}
%\frac{-ie}{\sqrt{2p_02p_0'2\omega'}}
%\int\, d^4x  M_{fi}(x)\,S{\rm e}^{i(k-p-p')x} \nonumber\\
=\frac{-ie}{\sqrt{2p_02p_0'2\omega'}}
\int\limits_\zeta^\infty dl
\, M_{fi}(l)(2\pi)^4\delta^4(k'+ lk -p-p'),
\label{III3}
\end{eqnarray}
where $k$, $k'$,  $p$ and $p'$ refer to the four-momenta of the
background (laser) field (\ref{III1}),
incoming probe photon, outgoing positron and electron,
respectively, the low limit  $\zeta$ is defined in Eq.~(\ref{I-zeta}).
The transition matrix $ M_{fi}(l)$
consists of four terms
\begin{eqnarray}
\, M_{fi}(l)=\sum\limits_{i=0}^3  M^{(i)}\,C^{(i)}(l)~,
\label{III4}
\end{eqnarray}
where transition matrices $M^{i}$ are determined by the Dirac structure
in the amplitude~(\ref{III3}) (cf.~\cite{TitovPEPAN}),
whereas the non-linear dynamics of
pair production is determined by the functions $C^{i}(l)$ expressed trough the
basic functions $Y_l$, $X_l$ which are an analog of the Bessel functions
\begin{eqnarray}
&&\hspace{-4 mm}C^{(0)}(l)=\widetilde Y_l(z){\rm e}^{il\phi_0},\,\,
\widetilde Y_l(z)=\frac{z}{2l} \left(Y_{l+1}(z) +
Y_{l-1}(z)\right) - \xi^2\frac{u}{u_l}\,X_l(z)'\,\,
C^{(1)}(l)=X_l(z)\,{\rm e}^{il\phi_0},\nonumber\\
&&\hspace{-4 mm}C^{(2)}(l)=\frac{1}{2}\left( Y_{l+1}{\rm e}^{i(l+1)\phi_0}
+ Y_{l-1}{\rm e}^{i(l-1)\phi_0}\right),\,\,
C^{(3)}(l)=\frac{1}{2i}\left( Y_{l+1}{\rm e}^{i(l+1)\phi_0}
- Y_{l-1}{\rm e}^{i(l-1)\phi_0}\right) %~,\nonumber\\
%C^{(0)}(l)&=&\widetilde Y_l(z){\rm e}^{il\phi_0}, \,\,
%\widetilde Y_l(z)=\frac{z}{2l} \left(Y_{l+1}(z) +
%Y_{l-1}(z)\right) - \xi^2\frac{u}{u_l}\,X_l(z)
~ \label{III25}
\end{eqnarray}
with
\begin{eqnarray}
&&\hspace{-4 mm}Y_l(z)=\frac{1}{2\pi} {\rm e}^{-il\phi_0}\int\limits_{-\infty}^{\infty}\,
d\phi\,{f}(\phi)
\,{\rm e}^{il\phi-i{\cal P}(\phi)} ~,\,\,
X_l(z)=\frac{1}{2\pi}{\rm e}^{-il\phi_0} \int\limits_{-\infty}^{\infty}\,
d\phi\,{f^2}(\phi)
\,{\rm e}^{il\phi-i{\cal P}(\phi)}~,\
\label{YX}\\
&&\hspace{-4 mm}{\cal P(\phi)}=z\int\limits_{-\infty}^{\phi}\,d\phi'\,
\cos(\phi'-\phi_0+\tilde\phi)f(\phi')
-\xi^2\zeta u\int\limits_{-\infty}^\phi\,d\phi'\,f^2(\phi')~.
\label{CP2}
\end{eqnarray}
The quantity $z$ is related to $\xi$, $l$, and
$u\equiv(k'\cdot k)^2/\left(4(k\cdot p)(k\cdot p')\right)$ via
$z=2l\xi\sqrt{\frac{u}{u_l}\left(1-\frac{u}{u_l}\right)}$;
with $u_l\equiv l/\zeta$.
The phase $\phi_0$ is equal to the azimuthal angle of
the direction of flight of the outgoing electron
in the $\ee$ pair rest frame $\phi_0=\phi_{p'}\equiv\phi_{e}$.
The quantity
\begin{eqnarray}
\zeta=\frac{4M_e^2}{s}
\label{zeta}
\end{eqnarray}
is important variable for the generalized Breit-Wheeler process:
$\zeta>1$ or $\zeta<1$  correspond to the above- and subthreshold-
pair production, respectively. The later one is mainly described  by
by the multi-photon interactions.

The production probability is presented as the integral over the variables $\phi_e$, $u$ and
$l$
\begin{eqnarray}
W=\frac{\alpha m\zeta^{1/2}}{16\pi N_0}
\int\limits_0^{2\pi_e} d\phi_e
 \int\limits_1^{\frac{l}{\zeta}}\frac{du}{u^{3/2}\sqrt{u-1}}
 \,\int\limits_\zeta^{\infty}\,dl\ w{(l)}
\label{III9}
\end{eqnarray}
with $N_0\simeq N$ and the partial probability
\begin{eqnarray}
w(l)=
 2 \widetilde Y^2_l(z) +\xi^2(2u-1)
\left(Y^2_{l-1}(z)+ Y^2_{l+1}(z)-2\widetilde
Y_l(z)X^*_l(z)\right)~. \label{III26-0}
\end{eqnarray}

\subsection{Pair production at small field intensities ($\xi^2\ll1$) }

In case of small $\xi^2\ll1$, implying  $z<1$,
we decompose $l=n +\epsilon$, where $n$ is the integer part of
$l$, yielding
\begin{eqnarray}
Y_l&\simeq&\frac{1}{2\pi}\int\limits_{-\infty}^{\infty}\,d\psi
\,{\rm e}^{il\psi -iz\sin\psi\,f(\psi+\phi_0)} f(\psi+\phi_0)\nonumber\\
&=& \frac{1}{2\pi}\int\limits_{-\infty}^{\infty}\,d\psi
\sum\limits_{m=0}^{\infty} \frac{(iz)^m}{m!}\sin^m\psi \,{\rm
e}^{i(n+\epsilon)\psi} f^{m+1}(\psi+\phi_0)~. \label{B5}
\end{eqnarray}
Similarly, for the function $X_l(z)$ the substitution $f^{m+1}\to
f^{m+2}$ applies. The dominant contribution to the integral
in (\ref{B5}) with
rapidly oscillating integrand comes from the term with $m=n$,
which results in
\begin{eqnarray}
Y_{n+\epsilon}\simeq \frac{z^n}{2^nn!}\,{\rm e}^{-i\epsilon\phi_0}
F^{(n+1)}(\epsilon)~,\qquad X_{n+\epsilon}\simeq
\frac{z^n}{2^nn!}\,{\rm e}^{-i\epsilon\phi_0}
F^{(n+2)}(\epsilon)~, \label{B6}
\end{eqnarray}
where the function $F^{(n)}(\epsilon)$ is the Fourier transform of
the function $f^n(\psi)$.

\begin{figure}[ht]
\begin{center}
\includegraphics[width=0.35\columnwidth]{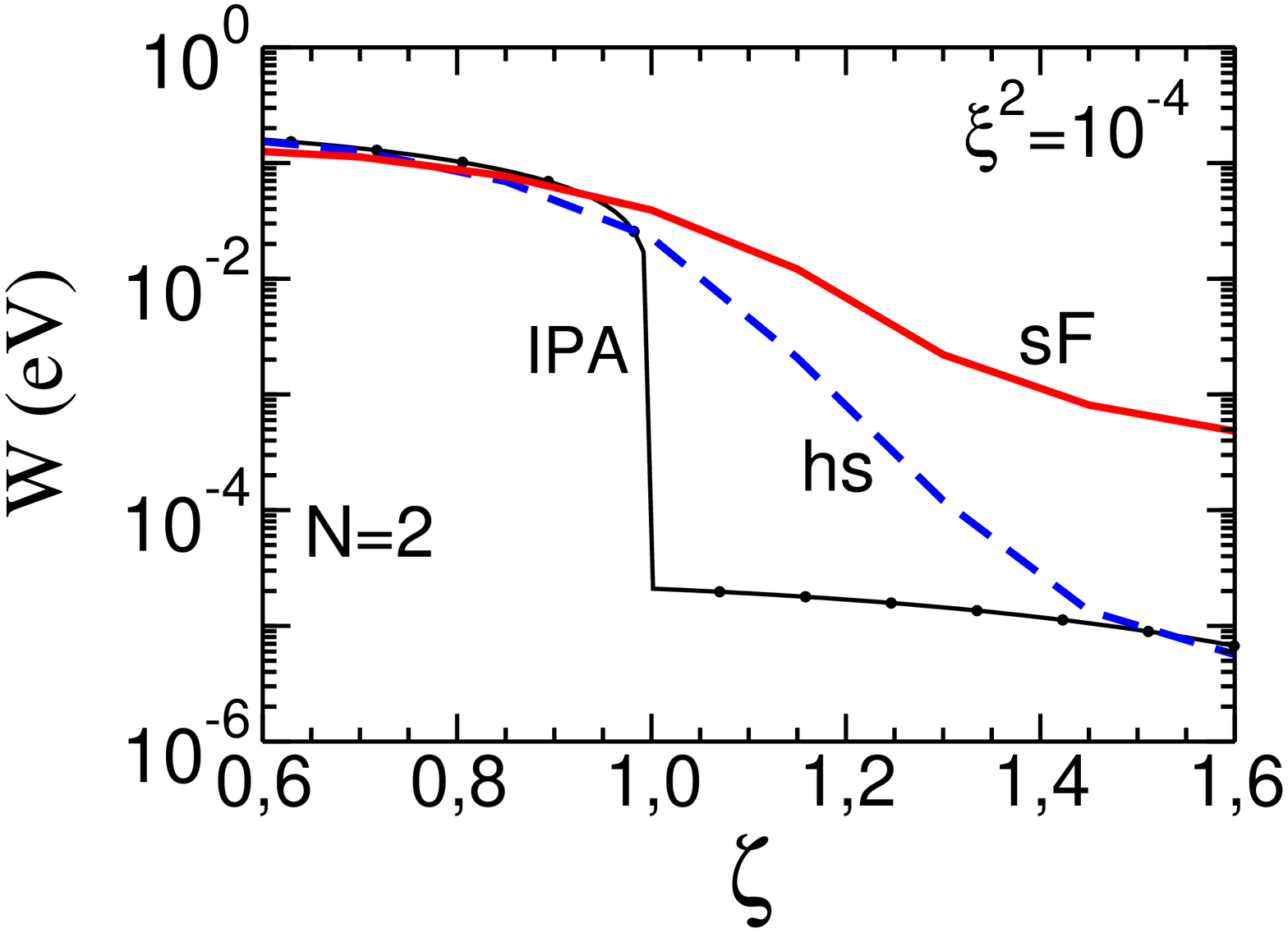}\qquad
\includegraphics[width=0.35\columnwidth]{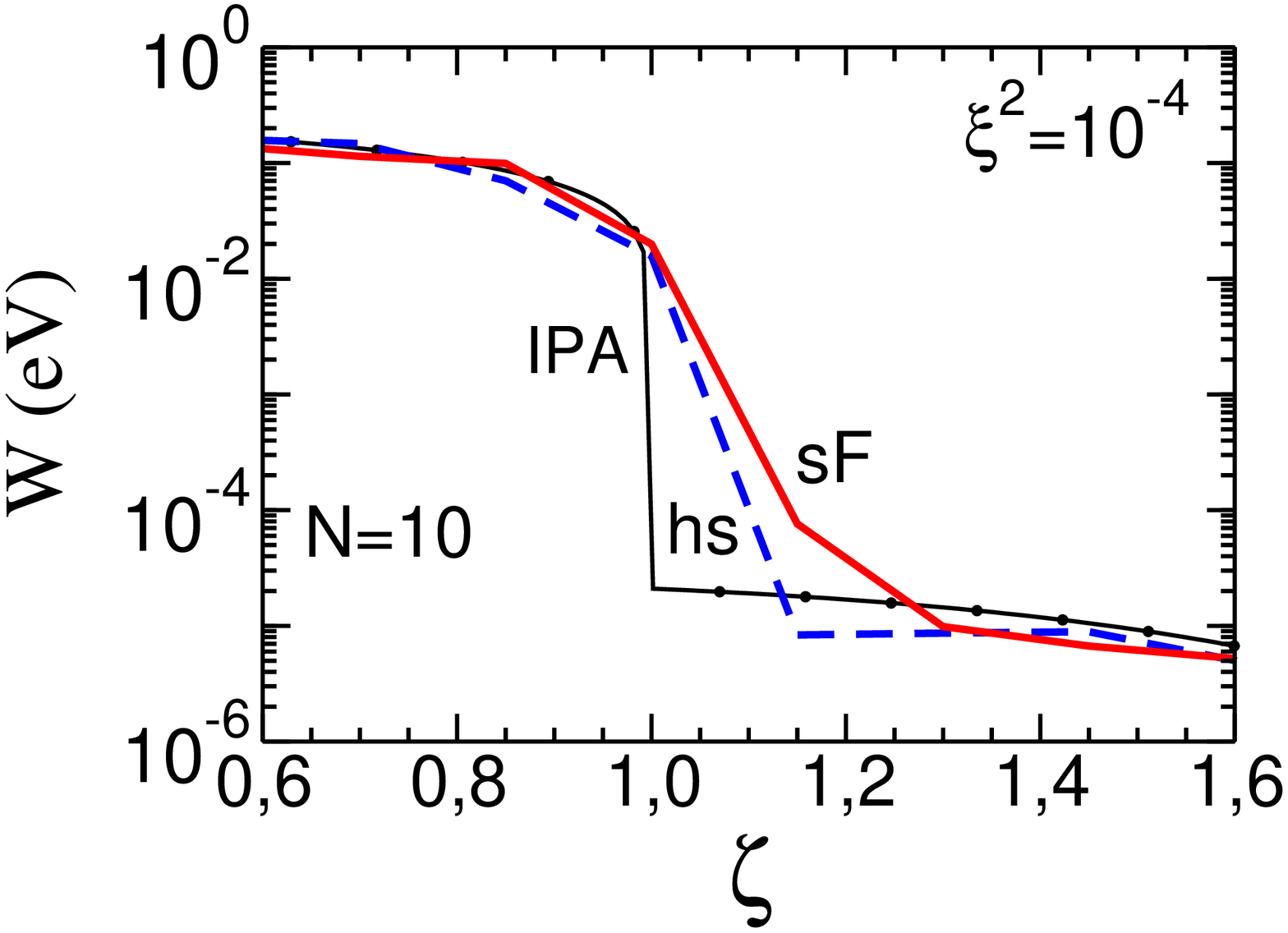}
\end{center}
\caption{\small{The total probability $W$ of the
 $\ee$ pair production as a function of $\zeta$ for short pulses
 with $\Delta=\pi N$ for $N=2$,\ and 10 shown in the left
 and right panels, respectively; $\xi^2=10^{-4}$.
 The dashed and solid curves correspond to the hyperbolic secant
 and symmetrized Fermi envelope shapes with $b/Delta=0.1$, respectively.
 The thin solid curves marked by dots depict the IPA
 result.
 \label{Fig:11} }}
\end{figure}

As an example, let us analyze the $\ee$ production near the
threshold, i.e. $\zeta\sim1$. In this case, the contribution with
$n=1$ is dominant and, therefore, the functions  $Y_{0+\epsilon}$
are crucial, including the first term in (\ref{III26-0}). The
functions $X_{0 +\epsilon}$ are not important because they are
multiplied  by the small $\xi^2$ and may be omitted. Negative
$\epsilon=\zeta-1$ and positive $\epsilon$ correspond to the
above- and sub-threshold pair production, respectively. The
function $Y_{0+\epsilon}$ reads
$Y_{0+\epsilon}=F^{(1)}(\epsilon)\,\exp[-i\phi_0\epsilon]$,
where
the Fourier transforms $F^{(1)}(x)$ for the hs and sF envelope
%\begin{eqnarray}
%
%F_{\rm hs}(l)&=&\frac{\Delta} {  2\cosh  {\frac12\pi\Delta  l}} ~,\nonumber\\
%
%F_{\rm sF}(l)&=&\frac{1+{\exp}\left[{-\frac{\Delta}{b}}\right] }
%{1-{\exp}\left[{-\frac{\Delta}{b}}\right] }\,
%\frac{ b\,\sin\Delta l } {\sinh \pi bl}~.
%
 %\frac{ {\exp}\left[{-\pi b l}\right] }
 %{ 1 - {\exp}\left[{-2\pi b l}\right] }~.
%\label{U5}
%\end{eqnarray}
decrease as a function of $l$ in a different way
$F_{\rm hs}(l)\sim\exp[-\frac{1}{2}\Delta l]$ and $F_{\rm sF}(l)\sim [-\pi bl] $
which is manifested in the spectra of $\ee$ pair production.
The $\phi_0$ dependence of the
production probability disappears in this case because the latter
one is determined by the quadratic terms of the $Y$ functions.
As we have seen the Fourier transform of the
envelope function plays important role in shape and absolute value
of the production probability.
As an example, in Fig.~\ref{Fig:11} we show the  total probability $W$ of $\ee$ emission
as a function of the
sub-threshold parameter $\zeta$ in the vicinity $\zeta\sim 1$.
 The dashed and solid curves correspond to the hyperbolic secant
 and symmetrized Fermi envelope shapes, respectively.
 The left and right panels correspond to the  short pulses
 with $\Delta=\pi N$ for $N=2$,\ and 10, respectively,
 at $\xi^2=10^{-4}$.
 For comparison, we present also the
 IPA results. Naturally, that in the above-threshold region, results of IPA and FPA
 are equal to each other. However, in the sub-threshold region, where $\zeta$
is close to integer numbers, the probability of FPA considerably exceeds (by more
than two orders of magnitude) the corresponding IPA
result. In the case of the hyperbolic secant envelope function, the
probability increases with decreasing  pulse duration. The results
of FPA and IPA become comparable at $N\geq 10$. Qualitatively,
this result is also valid for the case of the symmetrized Fermi
distribution. However, in this case the enhancement of the
probability in FPA is much greater. Other important details
may be found in~\cite{TitovPRA}.
%
%%%%%%%%%%%%%%%%%%%%%%%%%%%%%%%%%%%%%%%
\subsection{Effect of the finite carrier phase}

\begin{figure}[t]\qquad\qquad\quad\includegraphics[width=0.35\columnwidth]{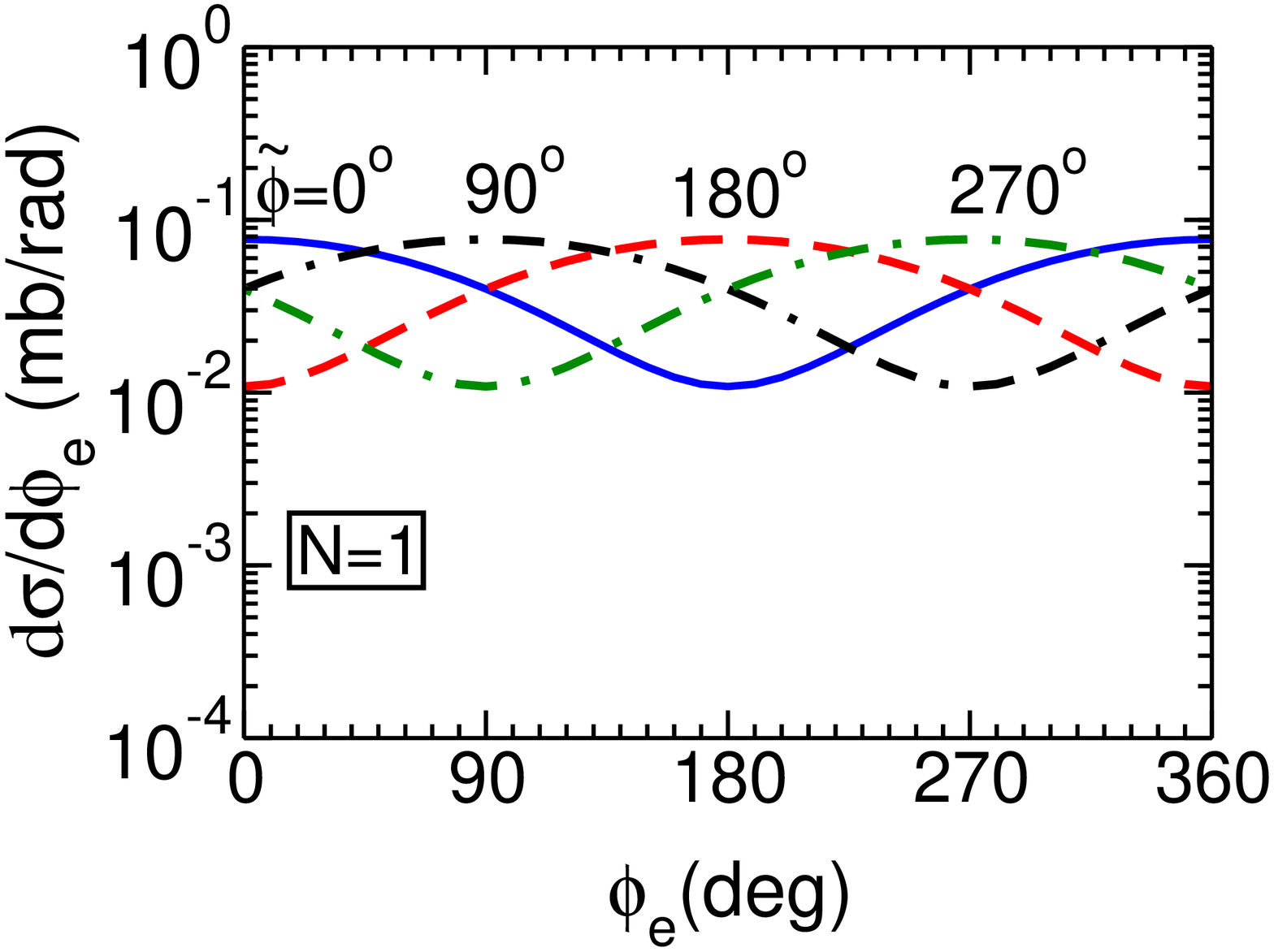}\qquad
\includegraphics[width=0.35\columnwidth]{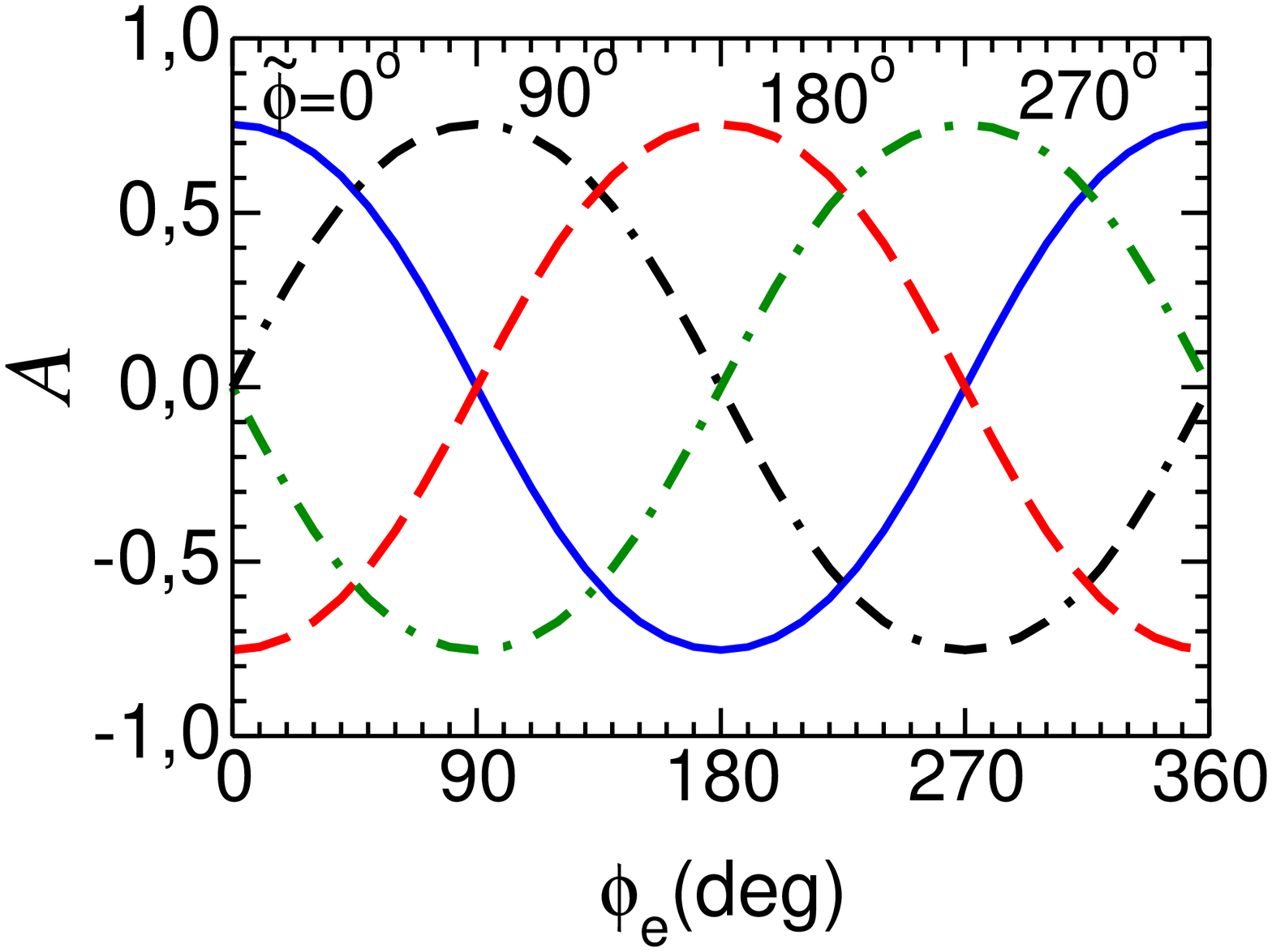}
\caption{\small{(Color online)
Left column: The differential cross section as a function of the azimuthal
angle of the direction of flight of the outgoing electron
$\phi_e$, for different values of the
carrier phase $\tilde\phi$ and for $N=1$.
The solid, dash-dash-dotted, dashed and dash-dotted
curves are for the CEP equal to 0,
90, 180 and 270 degrees, respectively.
Right column: The anisotropy~(\ref{U9}) for different
values of $\tilde\phi$ as in left column.
For $\xi^2=0.5$ and $ \zeta=4$.
 \label{Fig:CE1}}}
\end{figure}

 It is naturally to expect that the  effect of the finite carrier
 phase and appears in the azimuthal angle distribution of
 the outgoing electron (positron) in case of finite $\xi\sim1$ and
 smooth envelope function with $N<1$,
 because at this conditions the functions
 $C^{i}$ are greatly enhanced~\cite{TitovCF}.
  As an example, in Fig.~\ref{Fig:CE1} (left panels) we show
  the differential cross section $d\sigma/d\phi_e$
 of $\ee$ pair production as a function of the azimuthal angle
 $\phi_e$ for different values of the carrier envelope phase $\tilde\phi$
 and for  pulse durations
 $\Delta=N\pi$ with $N=$ 1, and for $\xi^2=0.5$.
 The calculation is done for the essentially multi-photon region with
 $\zeta=4$.
 The corresponding anisotropy of the electron (positron) emission defined
 as
  \begin{eqnarray}
{\cal A}=\frac{d\sigma(\phi_e) - d\sigma(\phi_e+\pi)}
{d\sigma(\phi_e) + d\sigma(\phi_e+\pi)}~,
\label{U9}
\end{eqnarray}
  are exhibited in
  Fig.~\ref{Fig:CE1} (right panels). One can see a strong dependence
  of the anisotropy as a function of CEP.
  The increase of the pulse duration leads
  to a decrease of the bump structure inn the differential cross sections
  and in absolute value of ${\cal A}$ and leads to the disappearance
  of the carrier phase effect.

\subsection{Pair production at large field intensity $(\xi^2\gg1)$}

At large values of $\xi^2\gg1$,
the basic functions $Y_l$ and $X_l$ in
Eq.~(\ref{YX}) can be expressed as follows
\begin{eqnarray}
{Y}_l=
%\frac{1}{2\pi}\int\limits_{-\infty}^{\infty}dl\,{\rm e}^{il\phi}\,f(\phi)\,g(\phi)
\int\limits_{-\infty}^{\infty}dq\,F^{(1)}(q)\,G(l-q)~,\qquad
{X}_l=\int\limits_{-\infty}^{\infty}dq\,F^{(2)}(q)\,G(l-q)~,
\label{H1}
\end{eqnarray}
where $F^{(1)}(q)$ and $F^{(2)}(q)$ are Fourier transforms of
the functions $f(\phi)$ and $f^{2}(\phi)$, respectively,
and $G(l)$ may be written as
\begin{eqnarray}
G(l)=\frac{1}{2\pi}\int\limits_{-\infty}^{\infty}d\phi
\,{\rm e}^{i\left( l\phi -z\sin\phi +\xi^2\zeta u\phi
\right)}~.
\label{H2}
\end{eqnarray}
In deriving this equation we have considered the following facts:
(i) at large $\xi^2$ the probability is isotropic, therefore we put
$\phi_0=0$, (ii) the dominant contribution to the rapidly oscillating
exponent comes from the region $\phi\simeq0$, where the difference of two
large values $l\phi$ and $z\sin\phi$ is minimal, and therefore,
one can decompose the last term in the function
${\cal P}(\phi)$ in (\ref{III21}) around $\phi=0$, and
(iii) replace in exponent $f(\phi)$ by $f(0)=1$.

Equation~(\ref{H2}) represent an asymptotic form of the Bessel
functions $J_{\tilde l}(z)$~\cite{WatsonBook}
with $\tilde l= l + \xi^2\zeta u$ at $\tilde l\gg 1$, $z\gg 1$,
and therefore the following identities
are valid
\begin{eqnarray}
G(\tilde l-1) - G(\tilde l+1)=2G_z'(\tilde l), \qquad
G(\tilde l-1) + G(\tilde l+1)=2\frac{\tilde l}{z} G(\tilde l)~,
\label{H3}
\end{eqnarray}
which allow to express the partial probability $w(\tilde l)$ in
(\ref{III26-0})
as a sum of the diagonal (relative to $\tilde l$) terms: $Y_{\tilde l}^2$,
$Y_{\tilde l}X_{\tilde l}$, $X_{\tilde l}^2$ and $Y^{'2}_{\tilde l}$.
The integral over $\tilde l$
from the diagonal term can be expressed as
\begin{eqnarray}
I_{YY}=\int\limits_{\tilde l_0}^{\infty}d{\tilde l}\,Y_l^2=
\int dq\,dq' F^{(1)}(q)\, F^{(1)}(q')
\int\limits_{\tilde l_0}^{\infty}d\tilde l
G(\tilde l-q)G(\tilde l-q')~,
\label{H4}
\end{eqnarray}
where $\tilde l_0=\zeta(1+\xi^2u$.
Taking into account that for the rapidly oscillating $G$ functions
$G(l-q)G(l-q')\simeq \delta(q-q')G^2(l-q)$ and
$\langle q \rangle\ll\langle l \rangle\ \sim \xi^2$ one gets
\begin{eqnarray}
I_{YY}=
\frac{1}{2\pi}\int\limits_{-\infty}^{\infty}d\phi f^{2}(\phi)
\int\limits_{\tilde l_0}^{\infty}d\tilde l G^2(\tilde l)
=N_{YY}\int\limits_{\tilde l_0}^{\infty}d\tilde l G^2(\tilde l)~.
\label{H44}
\end{eqnarray}
Similar expressions are valid
for the other diagonal terms with own normalization factors.
For  the $X^2_{\tilde l}$ term it is
$N_{XX}=\frac{1}{2\pi}\int\limits_{-\infty}^{\infty}d\phi
f^{4}(\phi)$, and for $Y_{\tilde l}X_{\tilde l}$,  $N_{YX}=\frac{1}{2\pi}\int\limits_{-\infty}^{\infty}d\phi
f^{3}(\phi)$. At large $\xi^2$,
the probability does not depend on the
 envelope shape, because only the central part of
the envelope is important. Therefore, for simplicity,
we choose the flat-top shape
with $N_{YY}=N_{YX}=N_{XX}=N_{0}=\Delta/\pi$ which is valid for any
smooth (at $\phi\simeq 0$) envelopes.

Making a change of the variable $l\to \tilde l= l+\xi^2\zeta u$
the variable $z$ takes the following form
\begin{eqnarray}
z^2=4\xi^2\zeta^2\left(uu_l -u^2\right)
=\frac{4\xi^2l_0^2}{1+\xi^2}\left(uu_{\tilde l} -u^2\right)
\label{H5}
\end{eqnarray}
with $l_0=\zeta(1+\xi^2)$ and $u_{\tilde l}\equiv {\tilde l}/{l_0}$,
that is exactly the same as the variable $z$ in IPA with the substitution
$l \to \tilde l$.
All these transformations
allow to express the total probability in a form similar to the
probability in IPA for large values of $\xi^2$ and a large number of
partial harmonics $n$, replacing the sum over $n$ by an integral
over $n$~\cite{Ritus-79}
\begin{eqnarray}
W&=&\frac12 {\alpha M_e\zeta^{1/2}}
\int\limits_{l_0}^{\infty}d\tilde l
\int\limits_1^{u_{\tilde l}}\frac{du}{u^{3/2}\sqrt{u-1}}
\{J^2_{\tilde l}(z)\nonumber\\
&+&\xi^2(2u-1)[
(\frac{{\tilde l}^2}{z^2} -1  ) {J}^2_{\tilde l}(z) + { J'}^2_{\tilde l}(z)
]\}~.
\label{H6}
\end{eqnarray}

Utilizing Watson's representation~\cite{WatsonBook}
for the Bessel functions at $\tilde l,\,z\gg1$ and
$\tilde l>z$,
$J_{\tilde l}(z)=({{2\pi\tilde l\tanh\alpha}})^{-1/2}
\exp[-\tilde l(\alpha -\tanh\alpha)]$ with
$\cosh\alpha={\tilde l}/{z}$,
and employing a saddle point approximation in the integration in
(\ref{H6}) we find the total
probability of $\ee$ production as (for details see Appendix~A of~\cite{TitovPEPAN})
\begin{eqnarray}
W=\frac{3}{8}\sqrt{\frac32} \frac{\alpha M_e \xi}{\zeta^{1/2}}
\,d\,
\exp\left[ -\frac{4\zeta}{3\xi}(1-\frac{1}{15\xi^2}) \right],
\,\,d=1+ \frac{\xi}{6\zeta}\left(1+\frac{\xi}{8\zeta} \right)~.
\label{H7}
\end{eqnarray}
This expression resembles the production probability in IPA which
is the consequence of the fact that, at $\xi^2\gg1$
in a short pulse, only the central
part of the envelope at $\phi\simeq 0$ is important.
In case of $\xi/\zeta<<1$, approximating
$d=1+{\cal O}(\xi/\zeta)$, the leading order
term recovers the Ritus result~\cite{Ritus-79}.

\begin{figure}[ht]
\begin{center}
\includegraphics[width=0.35\columnwidth]{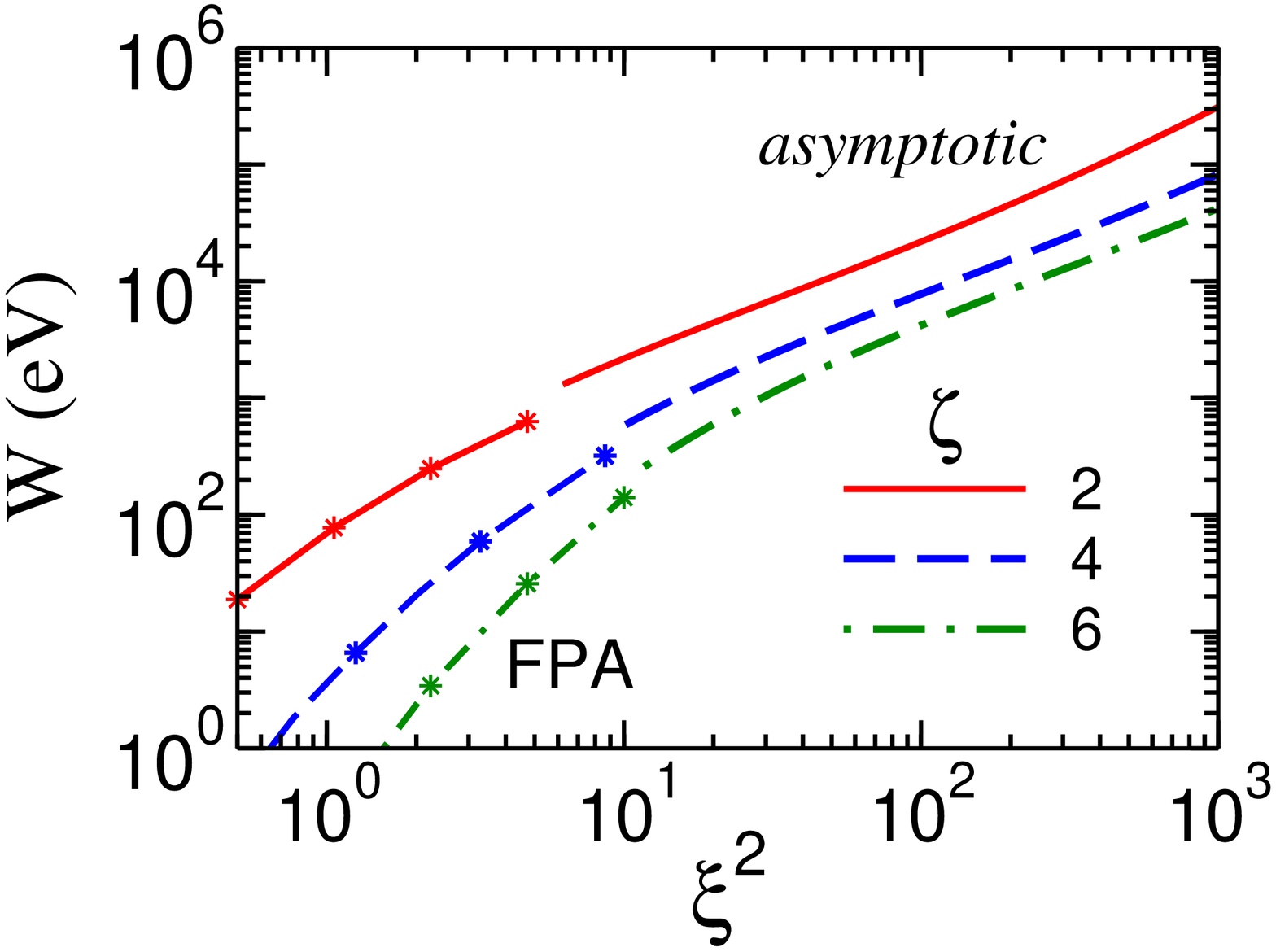}\qquad
\includegraphics[width=0.35\columnwidth]{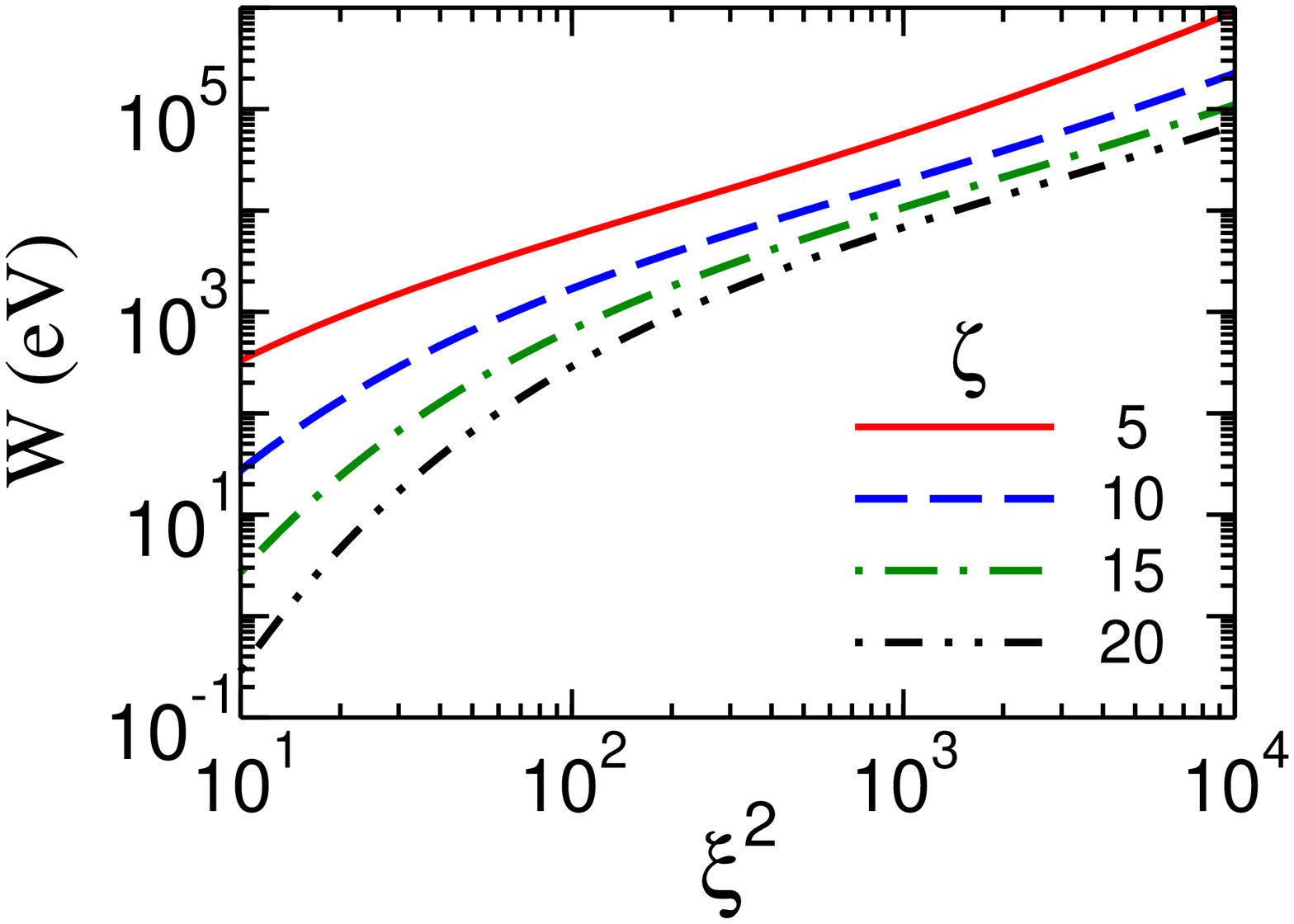}
\end{center}
\caption{\small{The total probability $W$ of the
$\ee$ pair production as a function of $\xi^2$
for various values of $\zeta$.
Left panel: Results of full numerical calculation in FPA
for finite values of $\xi^2\leq 10$
(curves marked by "stars"\ in "FPA"\ sections)
and  the asymptotic probability~(\ref{H7})
for large values of $\xi^2$ (sections labeled by "asymptotic")
at $\zeta=2$, 4 and 6.
Right panel: The asymptotic probability~(\ref{H7})
for various values of
$\zeta$ as indicated in the legend.
\label{Fig:15} }}
\end{figure}

For completeness, in Fig.~\ref{Fig:15} (left panel)
we present FPA results of a full numerical calculation
for finite values of $\xi^2\leq 10$
for the hyperbolic secant envelope shape with $N=2$
(curves are marked by "stars")
and the asymptotic probability calculated by Eq.~(\ref{H7})
at $\zeta=2$, 4 and 6, shown by solid, dashed and dot-dashed
curves, respectively.
The transition region between the two regimes
is in the neighborhood of $\xi^2\simeq 10$.
In the right panel, we show the production probability at
asymptotically large values of $\xi^2$
for $5 \leq \zeta\leq 20$. The exponential factor
in (\ref{H7}) is
most important at relatively low values of $\xi^2\sim 10$
(large  ${\zeta}/{\xi}$).
At extremely large values of $\xi^2$ (small ${\zeta}/{\xi}$ ), the
pre-exponential factor is dominant.

\section{Compton scattering in short laser pulse}

 The Compton scattering process, symbolically
 $e^-+ L\to e^{-}{}' + \gamma' $ is considered here
 as the spontaneous emission
 of one photon off an electron in an external e.m.\ field (\ref{III1}).
 Some important aspects of generalized Compton scattering
 were discussed elsewhere (for references see~\cite{TitovPEPAN}).
 Being crossing to the Breit-Wheeler $\ee$ pair production the
 structure of the matrix elements and cross sections (production
 rates) of the both processes are the the same. The principle difference
 between them is absent the threshold behaviour of both processes.
 Thus, in Breit-Wheeler $\gamma'+\gamma\to e^+ + e^-$ one has a
 minimum value of the energy  $\omega_{\rm min}(\gamma')$ of the probe
 photon $\gamma'$ responsible for two
 electron mass production (at fixed "target" photon energy $\omega(\gamma)$).
 The processes with subthreshold energy $\omega'<\omega_{\rm min}$
 or sub-threshold invariant variables $\zeta>1$ are determined by the multi-photon dynamics.
 The Compton process
 $e^-+ \gamma\to e^{-}{}' + \gamma'$ is always above threshold at any
 energy of incoming photon $\gamma$.
 Therefore extracting
 multi-photon interactions in such process is
 an incredibly difficult problem.

 In~\cite{TitovEPJD} we suggested  to use so-called
 partially integrated cross sections determined at fixed and large angle of
 outgoing photon $\theta'=170^0$
\begin{eqnarray}
 {\tilde\sigma_{}(\omega')} = \int\limits_{\omega'}^{\infty}
 d\bar\omega' \frac{d\sigma (\bar\omega')}{d\bar\omega'}
 =\int\limits_{l'}^{\infty} dl
 \frac{d\sigma(l)}{dl}~,
 \label{S6}
\end{eqnarray}
where ${d\sigma (\omega)}/{\omega}$ is the Compton scattering involving $l$ photons,
while the lower limit of integration $l'(\omega_{\rm min})$ ir related to the four momentum
of incoming electron $p(E,\mathbf{p})$ and laser frequency $\omega$
 \begin{eqnarray}
 l'=\frac{\omega'}{\omega}\,
 \frac{E+|\mathbf{p}|\cos\theta'}{E+|\mathbf{p}|-\omega'(1-\cos\theta')}~.
 \label{S66}
 \end{eqnarray}
 Experimentally, this can be realized by an absorptive medium
 which is transparent for frequencies above a certain threshold
 $\omega'$. Otherwise, such a partially integrated spectrum can be
 synthesized from a completely measured spectrum. Admittedly, the
 considered range of energies with a spectral distribution
 uncovering many decades is experimentally challenging.
 Thus the ratio $\omega'(l)/omega'(1)$ may be considered as a threshold parameter
 for the partly integrated Compron scattering.
 \begin{figure}[ht]
\begin{center}
\includegraphics[width=0.35\columnwidth]{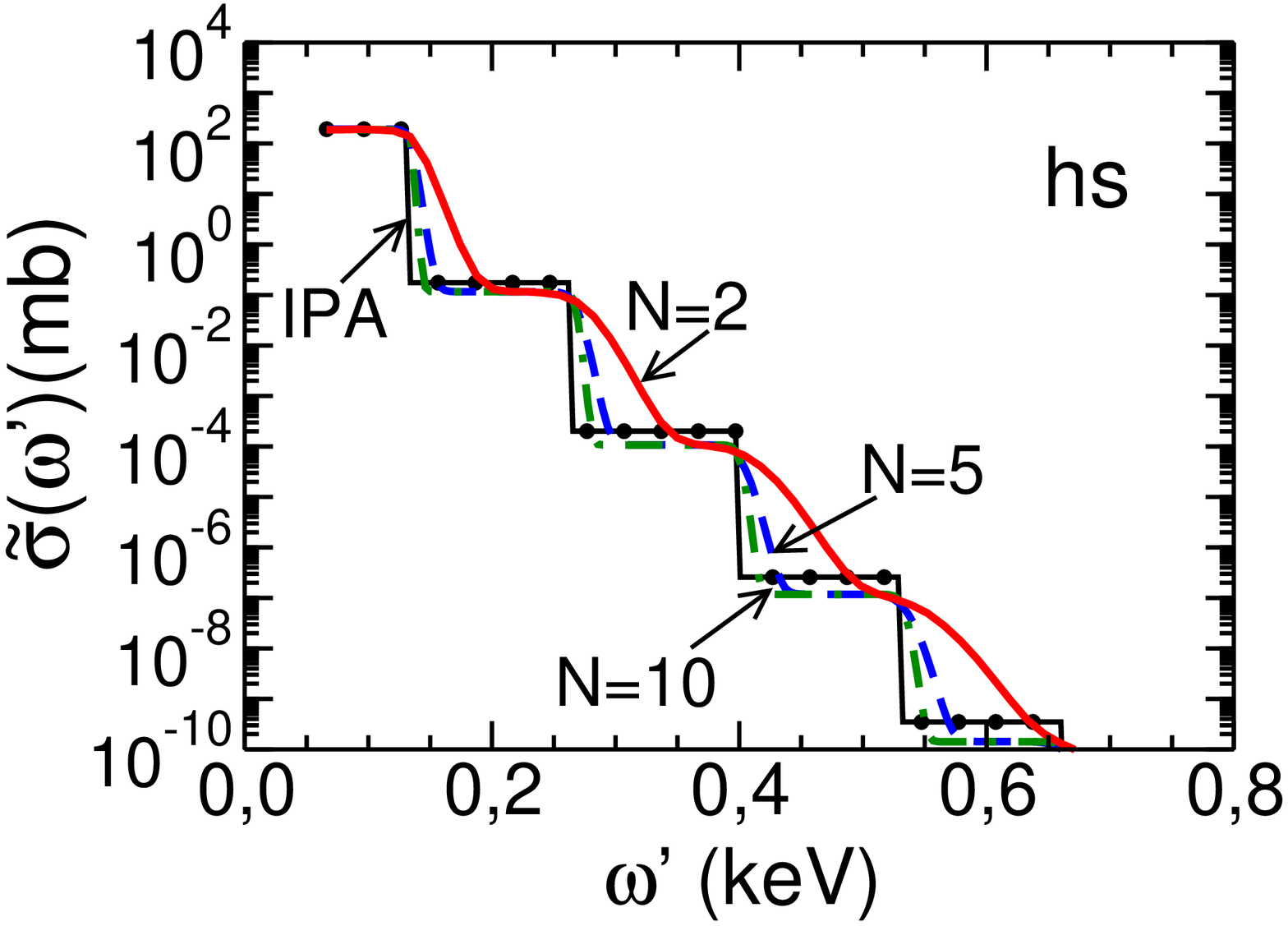}\qquad
\includegraphics[width=0.35\columnwidth]{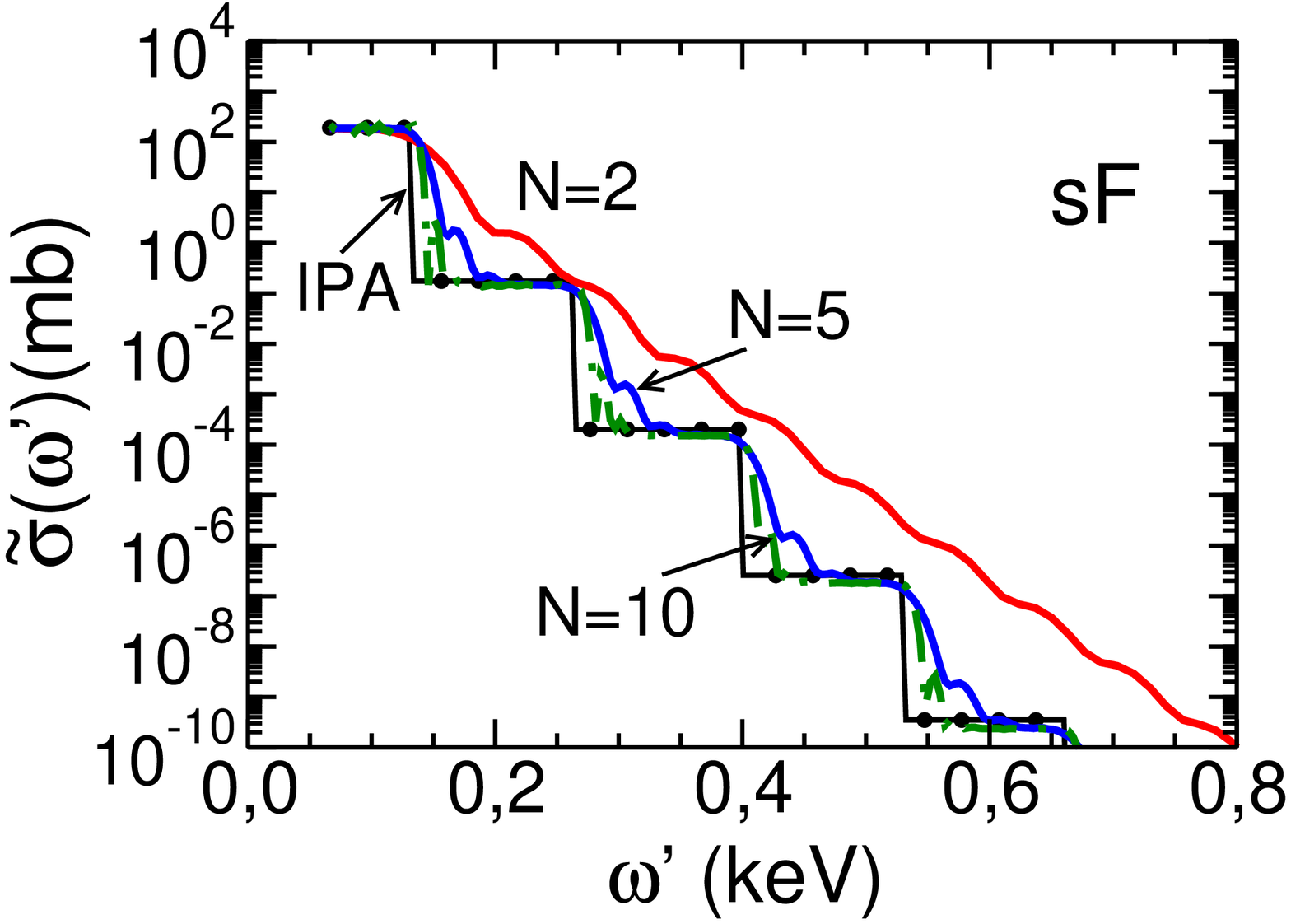}
\end{center}
 \caption{\small{The partially
 integrated cross section (\ref{S6})
 for $\xi^2=10^{-3}$.
 The thin solid curve marked by dots depicts the IPA result.
 The solid, dashed, and dot-dashed curves
 correspond to $N =2$, 5 and 10, respectively. Left and right
 panels are for hyperbolic secant (hs) and symmetrized Fermi (sF) envelopes.
% In latter the case, $b/\Delta=0.15$.
\label{gFig:7} }}
\end{figure}
  The partially integrated cross sections of Eq.~(\ref{S6})
  are presented in Fig.~\ref{gFig:7}.
  The thin solid curve (marked by dots) depicts results the photon
  emission in the infinite pulse (IPA) (cf.~\cite{TitovEPJD}).
 In this case the partially integrated cross section becomes a step-like
 function, where  each new step corresponds to the contribution of
 a new (higher) harmonic $n$,
 which can be interpreted  as $n$-laser photon process. Results for
 the finite pulse
 exhibited by solid, dashed,
 and dot-dashed curves
 correspond to $N=2, \,5$ and
 10, respectively. In the above-threshold region with
 $\omega'\leq \omega'_1$, the cross sections do not depend
 on the widths and shapes of the envelopes, and the results of IPA and FPA
 coincide. The situation changes significantly
 in the deep sub-threshold region, where $\omega'>\omega_1'$ $(l\gg1),\, n\gg1$.
 For  short pulses with $N\simeq 2$,
 the FPA results exceed that of IPA considerably,
 and the excess may reach
 several orders of magnitude, especially for the flat-top envelope
 shown by the solid curve in Fig.~\ref{gFig:7}~(right panel).
 However, when the number of oscillation in a pulse
 increases ($N\gtrsim 10$)
 there is  a qualitative convergence
 of FPA and IPA results, independently of the pulse shape.
 Thus, at $N=10$ and $\omega'=0.6$~keV the difference between predictions
 for hs and sF shapes is a factor of two, as compared with
 the difference of the few orders of
 magnitude at $N=2$ for the same value of $\omega'$.
\section{Summary}
 In summary, we briefly discussed main aspects of multi-photon dynamics
 in two important OCD processes in intensive laser field: Breit-Wheeler $\ee$ pair production
 and single photon radiation in propagation of an electron thought the laser beam.
 More detailed description of these and related topics  may be found in our review
 paper~\cite{TitovPEPAN}.
%%%%%%%%%%%%%%%%%%%%%%%%%%%%%%%%%%%%%%%%%%%%%%%%%%%%%%%%%%%%
\vspace{1cm}

%%%%%%%%%%%%%%%%%%%%%%%%%%%%%%%%%%%%%

%%%%%%%%%%%%%%%%%%%%%%%%%%%%

\end{document}